\documentclass{aastex}

\slugcomment{ApJL, in press}

\shorttitle{C$_{60}$}
\shortauthors{Sellgren et al.}

\begin{document}

\title{C$_{60}$ in Reflection Nebulae}

\author{Kris Sellgren\altaffilmark{1}}
\email{sellgren@astronomy.ohio-state.edu}

\author{Michael W. Werner\altaffilmark{2}}

\author{James G. Ingalls\altaffilmark{3}}

\author{J. D. T. Smith\altaffilmark{4}}

\author{T. M. Carleton\altaffilmark{5}}

\and

\author{Christine Joblin\altaffilmark{6,7}}

\altaffiltext{1}
{Department of Astronomy, Ohio State University, Columbus, OH 43235, USA}
\altaffiltext{2}
{Jet Propulsion Laboratory, California Institute of Technology, Pasadena, 
CA 91109, USA}
\altaffiltext{3}
{Spitzer Science Center, California Institute of Technology, 
Pasadena, CA 91125, USA}
\altaffiltext{4}
{Ritter Astrophysical Research Center, University of Toledo, 
Toledo, OH 43603, USA}
\altaffiltext{5}
{Steward Observatory, University of Arizona, Tucson, AZ 85721, USA}
\altaffiltext{6}
{Universit\'e de Toulouse, UPS, CESR, 9 ave colonel Roche, 
F-31028 Toulouse cedex 4, France}
\altaffiltext{7}
{CNRS, UMR 5187, 31028 Toulouse, France}

\begin{abstract}
The fullerene C$_{60}$ has four infrared-active vibrational transitions 
at 7.0, 8.5, 17.4 and 18.9 $\mu$m.  
We have previously observed emission features at 
17.4 and 18.9 $\mu$m in the reflection nebula NGC 7023 and 
demonstrated spatial correlations suggestive of a common origin. 
We now confirm our earlier identification of these features with 
C$_{60}$ by detecting a third emission feature at 
7.04 $\pm$ 0.05 $\mu$m in NGC 7023. 
We also report the detection of these three C$_{60}$ features in the 
reflection nebula NGC 2023.  
Our spectroscopic mapping of NGC 7023 shows that the 
18.9 $\mu$m C$_{60}$ feature peaks on the central star and
that the 16.4 $\mu$m emission feature due to 
polycyclic aromatic hydrocarbons  peaks 
between the star and a nearby photodissociation front.
The observed features in NGC 7023 are consistent with emission
from UV-excited gas-phase C$_{60}$.
We find that 0.1--0.6\% of interstellar carbon
is in C$_{60}$;
this abundance is 
consistent with those from previous upper limits and possible 
fullerene detections in the interstellar medium.
This is the first firm detection of neutral C$_{60}$ in the
interstellar medium.

\end{abstract}

\keywords{ISM: molecules --- ISM: lines and bands --- 
ISM: individual objects (NGC 7023, NGC 2023) --- line: identification}

\section{Introduction}

Fullerenes are cage-like molecules (spheroidal or ellipsoidal)
of pure carbon, such as C$_{60}$, C$_{70}$, C$_{76}$, and C$_{84}$.
C$_{60}$, also known as 
buckminsterfullerene, is the most stable fullerene and can account for
up to 50\% of the mass of fullerenes generated in the
laboratory
\citep{Kroto85}.
Theorists have suggested that fullerenes
might form around stars with carbon-rich atmospheres,
such as carbon stars, Wolf-Rayet (WC) stars,
and carbon-rich, hydrogen-poor R Cr B stars
\citep{Kroto92, Goeres92, Cherchneff00, Pascoli00}.
Fullerenes may also form as part of the carbon-rich grain
condensation process
known to occur in the ejecta of Type II supernovae
\citep{Clayton01}.
Hydrogenated
amorphous carbon grains in the interstellar medium (ISM)
may also decompose after interstellar shocks into
polycyclic aromatic hydrocarbons (PAHs)
and fullerenes 
\citep{Scott97};
PAHs comprise
9--18\% of
interstellar carbon 
\citep{Joblin92, Tielens08}.
Fullerenes might also form via cold interstellar gas-phase chemistry
\citep{Bettens96, Bettens97}.

\citet{Foing94} propose that
two diffuse interstellar bands at 958 and 963 nm are
due to singly ionized C$_{60}$, or C$_{60}^+$,
although the identification is debated
\citep{Maier94, Jenniskens97}.
\citet{Misawa09}
attribute additional diffuse interstellar bands at
902, 921, and 926 nm to C$_{60}^+$.
No fullerenes were found 
towards carbon-rich post-AGB stars
\citep{SB89, SS89},
carbon stars
\citep{Clayton95, Nucci05},
or R CrB stars 
\citep{Clayton95, Lambert01}.
\citet{Cami10} 
have recently detected C$_{60}$ and C$_{70}$ 
in the planetary nebula Tc 1 
(\object{IC 1266}).

Observational evidence for neutral fullerenes in
the ISM, however,
has been elusive to date. 
No neutral fullerenes have yet been found
in the diffuse ISM
\citep{SS89, Herbig00},
dense molecular clouds
\citep{Nucci05},
or at the photodissociation front in the reflection nebula 
\object{NGC 7023}
\citep{Moutou99}.

C$_{60}$ has four infrared-active vibrational transitions,
at 7.11, 8.55, 17.5, and 19.0 $\mu$m
at 1200 K
(gas-phase; \citealt{Frum91}), 
and at 6.98, 8.44, 17.3, and 18.9 $\mu$m at
2.4 K
(para-H$_2$ matrix isolation; \citealt{Sogoshi00}).
On this basis, we tentatively identified the 
17.4 and 18.9 $\mu$m ISM
emission features in the reflection nebula
NGC 7023 as due to C$_{60}$
\citep{WUS04, SUW07}.

We provide here additional evidence for the
presence of C$_{60}$ in reflection nebulae.
We report here the detection of the 
predicted C$_{60}$ feature at 7.04 $\pm$ 0.05 $\mu$m
in NGC 7023.
We also report the detection of C$_{60}$ features
at 7.04, 17.4 and 18.9 $\mu$m 
in a second reflection nebula,
\object{NGC 2023}.
The C$_{60}$ 8.5 $\mu$m 
feature is too blended with strong 8.6 $\mu$m 
PAH emission 
to be detected in reflection nebulae.
We have found in NGC 7023
\citep{SUW07}
that the 16.4 $\mu$m emission feature,
attributed to PAHs
\citep{Moutou00},
has a
spatial distribution distinct from 
that of the 18.9 $\mu$m emission feature.
We now compare the spatial distributions 
of the 16.4, 17.4, and 18.9 $\mu$m
emission features in NGC 7023, and
find additional support for the C$_{60}$ identification.
We also discuss the excitation mechanism for
the infrared emission of C$_{60}$.

\section{Observations}

We used the {\it Spitzer Space Telescope} 
\citep{Spitzer}
with the Infrared Spectrograph
(IRS; \citealt{Houck04}) to obtain 5--38 $\mu$m spectra of 
NGC 2023 
(PI Sellgren, pid 40276, aorkeys = 
\dataset[ADS/Sa.Spitzer#0023912704]{23912704}, 
\dataset[ADS/Sa.Spitzer#0023911168]{23911168}) 
and NGC 7023
(PI Sellgren, pid 40276, aorkeys = 
\dataset[ADS/Sa.Spitzer#0023911424]{23911424}, 
\dataset[ADS/Sa.Spitzer#0023911680]{23911680}). 
We obtained spectra with the short-wavelength low-resolution module SL 
(5--14 $\mu$m; $\lambda / \Delta\lambda$ = 60--120),
the long-wavelength low-resolution module LL
(14--38 $\mu$m; $\lambda / \Delta\lambda$ = 60--120),
and the short-wavelength high-resolution module SH 
(9.5--19.5 $\mu$m; $\lambda / \Delta\lambda$ = 600).
We chose nebular positions
(29\arcsec\ west, 8\arcsec\ south of HD 37903 in NGC 2023;
25\arcsec\ east, 4\arcsec\ north of HD 200775 in NGC 7023)
with a strong ratio
of the 18.9 $\mu$m feature relative
to the 16.4 $\mu$m PAH feature.
We used matched aperture extraction in CUBISM 
\citep{Smith07_cubism}
to extract SL, LL, and SH spectra 
in regions of overlap
between these spectral modules.
The extraction aperture was 
10.2\arcsec$\times$10.2\arcsec\ in NGC 2023
and 7.5\arcsec$\times$9.2\arcsec\ in NGC 7023.

We also retrieve from the Spitzer archive a spectral data cube for NGC 7023 with LL
(PI Joblin, pid 3512; aorkey =
\dataset[ADS/Sa.Spitzer#0011057920]{0011057920}). 
We use CUBISM to derive spectral images in
the 16.4, 17.4, and 18.9 $\mu$m features 
and 0--0 S(1) H$_2$ for NGC 7023.
For the spectrum of each spatial pixel, we define a local continuum surrounding 
an emission feature or line,
and subtract it before deriving the
feature or line intensity.

We search for bad pixels and correct them with CUBISM
before extracting final spectra.
We subtract dedicated sky spectra for the 5--38 $\mu$m
spectra of NGC 2023 and NGC 7023; no sky subtraction is
done for the spectral mapping.

\section{Results}

\begin{figure*}
\figurenum{1}
\includegraphics[scale=0.80]
{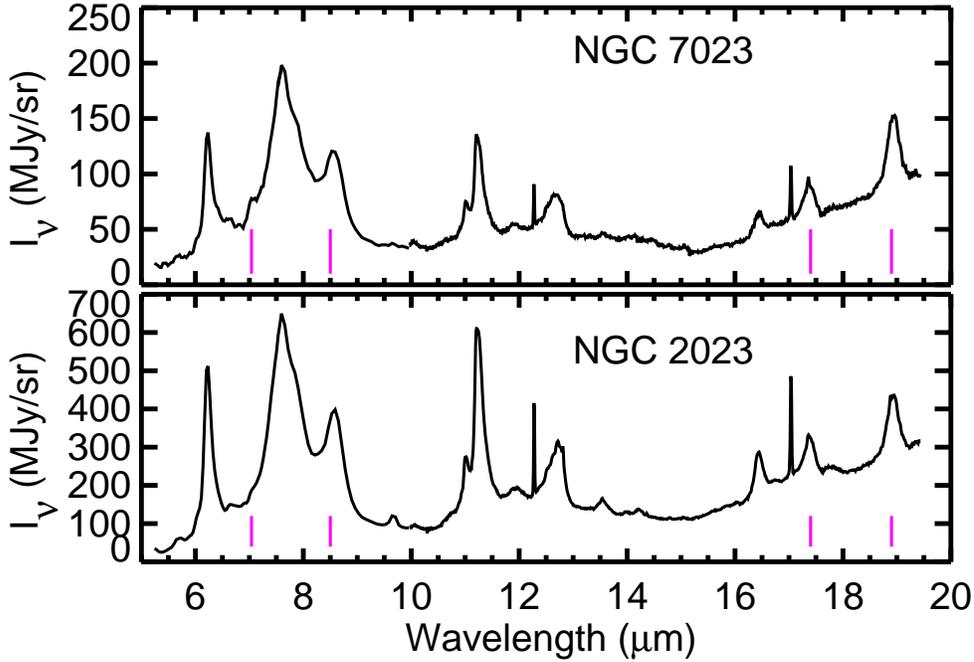}
\caption{
{\it Spitzer}-IRS spectra 
{\it (solid curves)} of NGC 7023 
(25\arcsec\ east, 4\arcsec\ north of HD 200775;
{\it top })
and NGC 2023 
(29\arcsec\ west, 8\arcsec\ south of HD 37903;
{\it bottom}),
obtained with 
the short-wavelength low resolution module (SL; 5.2--10.0 $\mu$m;
$\lambda / \Delta\lambda$ = 60--120)
and the short-wavelength high-resolution module (SH; 10.0--19.5 $\mu$m;
$\lambda / \Delta\lambda$ = 600).
We mark C$_{60}$ lines at 7.04, 8.5, 17.4 and 18.9 $\mu$m
{\it (vertical lines)}.
The strong emission feature at 8.6 $\mu$m is due to PAHs.
H$_2$ emission lines fall at
9.66, 12.3, and 17.0 $\mu$m.
\label{spectrum_sh}
}
\end{figure*}

Figure \ref{spectrum_sh} illustrates
our SL and SH spectra in NGC 2023 and NGC 7023.
The 17.4 and 18.9 $\mu$m emission features are prominent,
and coincident with C$_{60}$ wavelengths.

\begin{figure*}
\figurenum{2}
\includegraphics[angle=0,scale=0.90]
{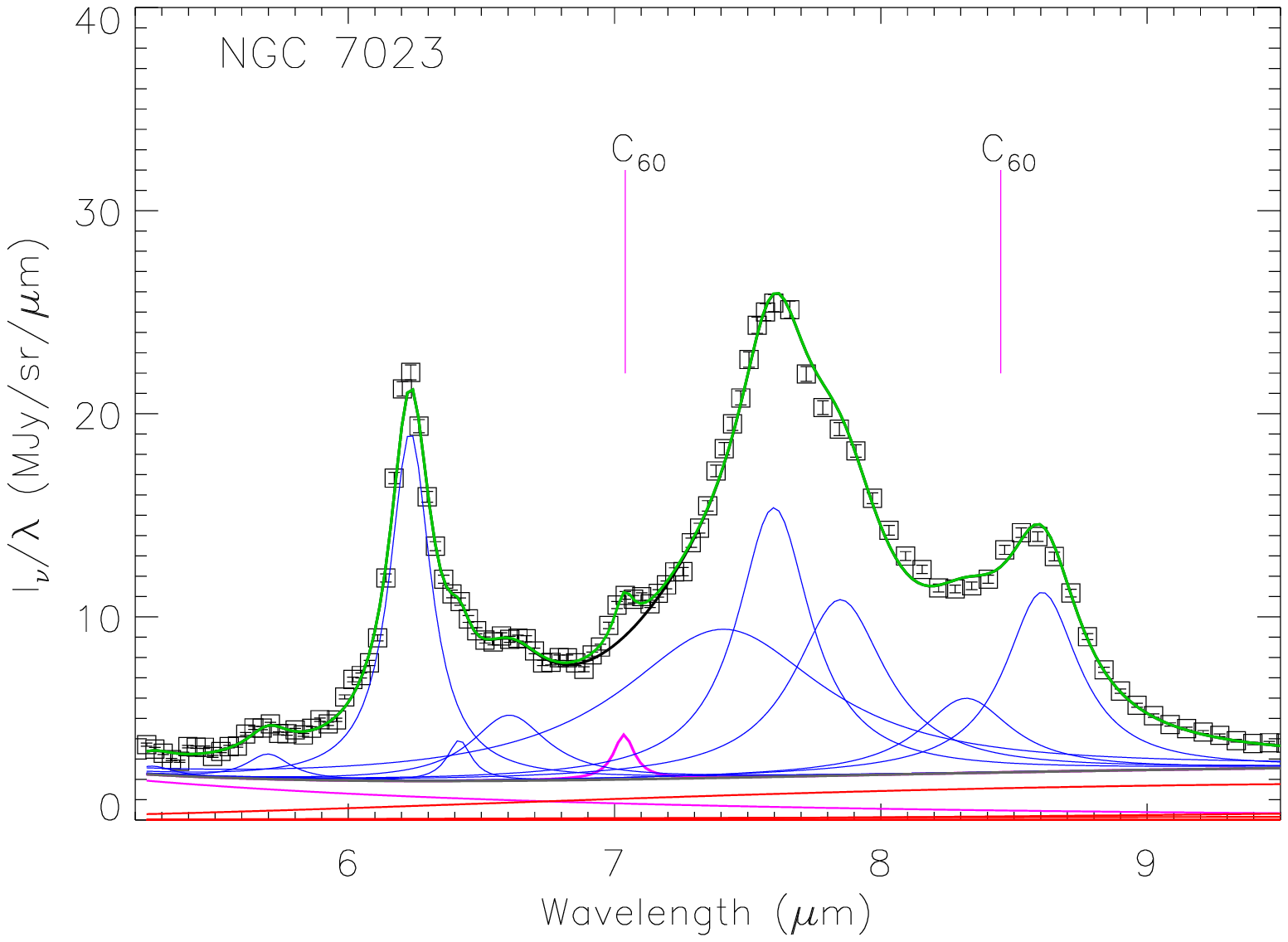}
\caption{
{\it Spitzer}-IRS 5--9 $\mu$m spectrum 
of NGC 7023
{\it (open squares)},
obtained with the short-wavelength low-resolution module (SL;
$\lambda / \Delta \lambda$ = 60--120).
We mark
C$_{60}$ lines at 7.04 and 8.5 $\mu$m 
{\it (vertical lines)}.
We show the individual contributions of PAH features at
5.3, 5.7, 6.2, 6.4, 6.7, 7.4, 7.6, 7.8, 8.3, and 8.6 $\mu$m
to the spectrum, by decomposing the spectrum with
PAHFIT \citep{Smith07_pahfit} and then overplotting the
Drude profile of each feature (blue curves).
The Drude fit to the C$_{60}$ feature we detect at 
7.04 $\pm$ 0.05 $\mu$m is 
highlighted (magenta curve).
The 8.5 $\mu$m C$_{60}$ feature
is blended with the strong 8.6 $\mu$m PAH feature.
\label{spectrum_sl}
}
\end{figure*}

We show the 5--9 $\mu$m SL spectrum of
NGC 7023 in Figure \ref{spectrum_sl}.
We clearly detect an emission feature at 7.04 $\pm$ 0.05 $\mu$m.
This feature is coincident, within the uncertainties, with the 
wavelength of the expected C$_{60}$ line.
We highlight this emission feature by 
using PAHFIT
\citep{Smith07_pahfit} 
to fit the 5--9 $\mu$m spectrum
with a blend of PAH emission features in addition to the
new emission feature at 7.04 $\mu$m.
The full-width at half-maximum 
of the 7.04 $\mu$m C$_{60}$ feature is 0.096 $\pm$ 0.012 $\mu$m,
significantly broader than our spectral resolution.
We also detect the 7.04 $\mu$m C$_{60}$ feature in
NGC 2023.
We present the C$_{60}$ band  intensities in Table \ref{table1}.

\begin{deluxetable}{lrrr}
\tablewidth{0pt}
\tablecaption{
Observed\tablenotemark{a}
and Calculated\tablenotemark{b}
C$_{60}$ Intensity 
Ratios\tablenotemark{c} 
}
\tablehead{    &
\colhead{$I_{7.04}$/$I_{18.9}$}  &
\colhead{$I_{8.5}$/$I_{18.9}$}  &
\colhead{$I_{17.4}$/$I_{18.9}$} 
   }
\startdata
{\bf Object} & & & \\
NGC 7023 ($\lambda$/$\Delta \lambda$=60--130)  & 
0.82 $\pm$ 0.12 & \nodata & 0.42 $\pm$ 0.02  \\
NGC 7023 ($\lambda$/$\Delta \lambda$=600)  & 
\nodata & \nodata & 0.33 $\pm$ 0.01 \\
NGC 2023 ($\lambda$/$\Delta \lambda$=60--130)  & 
0.086 $\pm$ 0.004 & \nodata & 0.47 $\pm$ 0.01  \\
{\bf Absorbed photon energy} & & & \\
5 eV  & 0.46--0.58 & 0.41--0.43 & 0.28--0.38  \\
10 eV  & 0.76--0.94 & 0.57--0.59 & 0.28--0.38 \\
15 eV  & 0.97--1.20 & 0.67--0.71 & 0.29--0.38  \\

\enddata

\tablenotetext{a}{Observed intensity ratios, derived using
PAHFIT \citep{Smith07_pahfit}.  
We give statistical uncertainties; systematic fitting uncertainties 
are 15\% for the 7.04 $\mu$m intensity ratio
and 30\% for the 17.4 $\mu$m intensity ratio.
The observed 17.4 $\mu$m feature has not been
corrected for PAH emission blended with it.}
\tablenotetext{b}{Emission spectrum calculated with Monte Carlo code
\citep{Joblin02} for molecular cooling cascade 
after absorbing a stellar photon.
C$_{60}$ vibrational data from
\citet{Menendez00}, \citet{Choi00}, and \citet{Schettino01}.}
\tablenotetext{c}{Intensities (W m$^{-2}$ sr $^{-1}$)
normalized to the 18.9 $\mu$m feature intensity.}
\label{table1}
\end{deluxetable}

\begin{figure*}
\figurenum{3}
\includegraphics[angle=0,scale=1.0]
{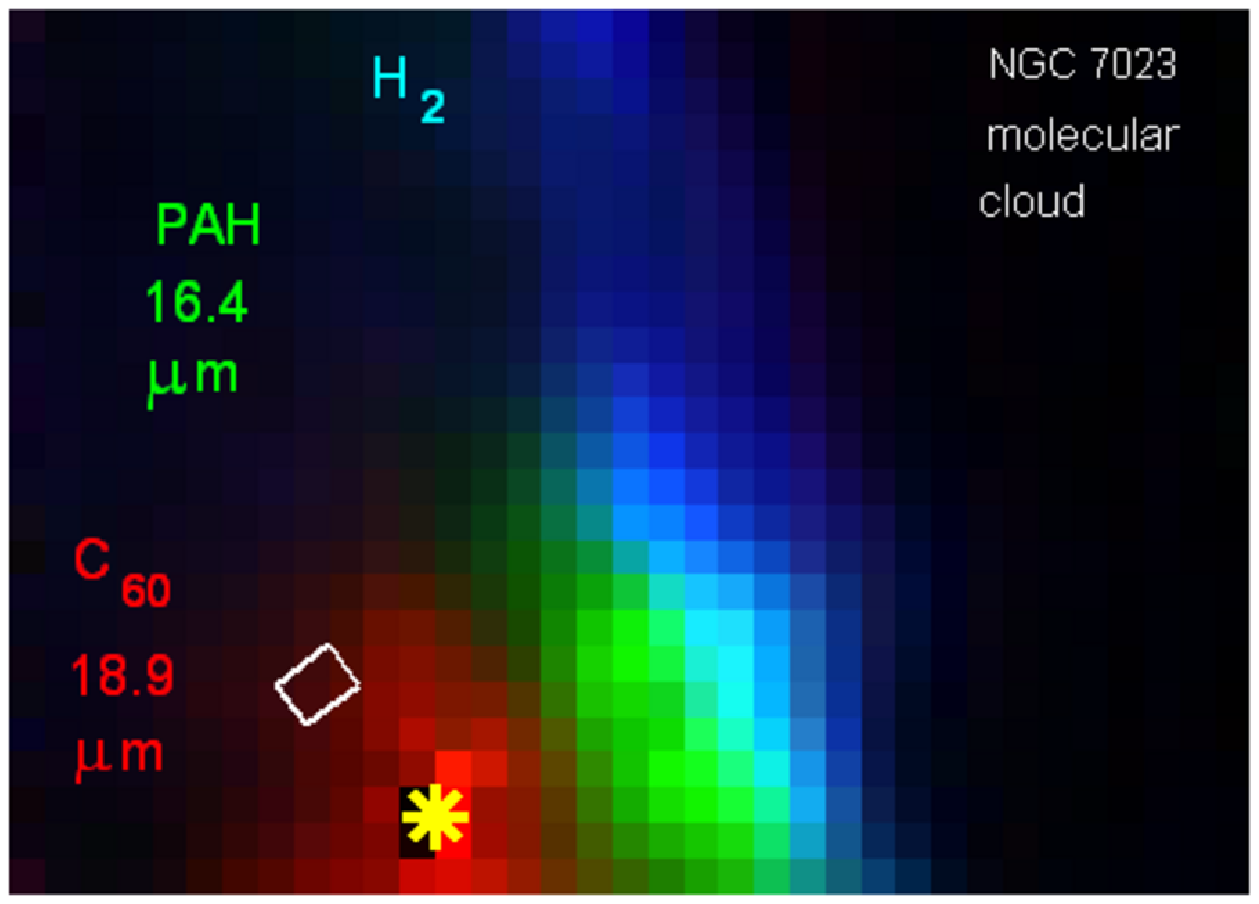}
\caption{
Three-color image of NGC 7023, in
18.9 $\mu$m C$_{60}$ emission {\it (red)},
16.4 $\mu$m PAH emission {\it (green)}, and
17.0 $\mu$m 0--0 S(1) H$_2$ emission {\it (blue)}.
We constructed each image by analyzing 
{\it Spitzer}/IRS-LL long-slit spectra with
CUBISM \citep{Smith07_cubism}.
We illustrate where we measured
the 5--38 $\mu$m spectra shown in Figs. 
\ref{spectrum_sh} and \ref{spectrum_sl} 
{\it (white rectangle)}.
We mark the location of HD 200775 {\it (star)}.
Each pixel is 5.1$''$ $\times$ 5.1$''$.
The 18.9 $\mu$m C$_{60}$ feature
peaks on the central star, 
while the 16.4 $\mu$m PAH
emission is brightest between the 18.9 $\mu$m
emission region and the photodissociation front
traced by H$_2$ emission.
\label{image1}
}
\end{figure*}

In our previous long-slit spectroscopic investigation of NGC 7023
\citep{SUW07}, we found that the 18.9 $\mu$m feature peaks
closer to the central star than PAHs.
We now illustrate this more clearly 
with the LL spectroscopic map extracted in NGC 7023
(Figure \ref{image1}).
The 18.9 $\mu$m emission is clearly centered on the star.
By contrast, the 16.4 $\mu$m PAH emission peaks outside the
region of maximum 18.9 $\mu$m emission, in a layer between the
star and the molecular cloud.
The photodissociation front at the UV-illuminated front surface of
the molecular cloud is delineated by 0--0 S(1) H$_2$ emission
at 17.0 $\mu$m.

\begin{figure*}
\figurenum{4}
\includegraphics[angle=0,scale=1.0]
{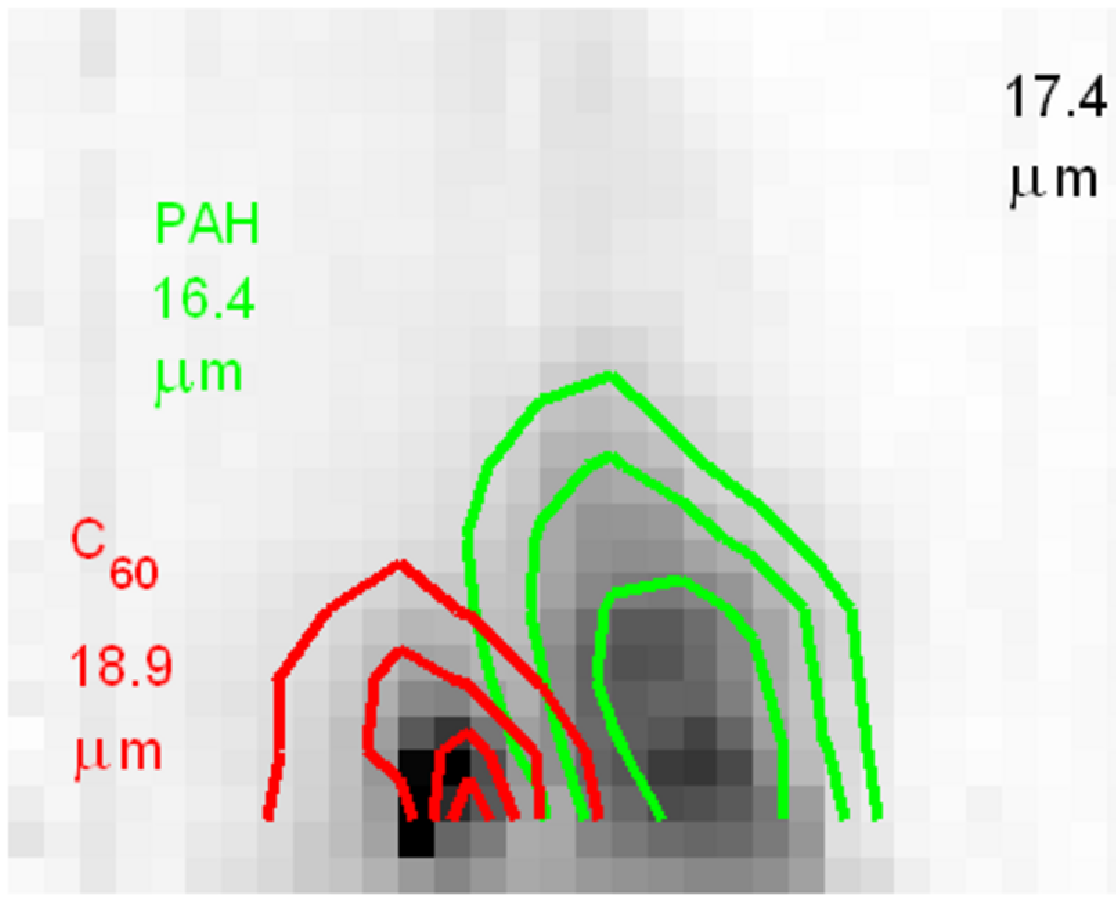}
\caption{
Image of 17.4 $\mu$m emission in NGC 7023
{\it (gray-scale image)},
derived from 
{\it Spitzer}/IRS-LL long-slit spectra analyzed with
CUBISM \citep{Smith07_cubism}.
We overlay contours of
18.9 $\mu$m C$_{60}$ emission {\it (red)}
and
16.4 $\mu$m PAH emission {\it (green)}.
Each pixel is 5.1$''$ $\times$ 5.1$''$.
One component of the 17.4 $\mu$m emission is
co-spatial with 
18.9 $\mu$m C$_{60}$ emission and
the other component of the 17.4 $\mu$m 
emission 
follows 16.4 $\mu$m PAH emission.
This illustrates that the 17.4 $\mu$m emission feature is
a blend of a PAH feature and 17.4 $\mu$m C$_{60}$ emission.
\label{image2}
}
\end{figure*}

Our previous observations \citep{SUW07}
suggested that the 17.4 $\mu$m feature might be a blend
of a PAH feature and an emission feature with the same spatial
distribution as the 18.9 $\mu$m feature.
We now confirm that this is the case with IRS/LL spectroscopic
imaging.
We show an image of the 17.4 $\mu$m emission from NGC 7023
in Figure \ref{image2}, overlaid with contours of 18.9 $\mu$m
and 16.4 $\mu$m emission.
The 17.4 $\mu$m emission clearly shows one peak on the central
star, coincident with 18.9 $\mu$m C$_{60}$ emission,
and a second peak co-spatial with 16.4 $\mu$m PAH emission.
Thus there is an ISM component with emission
features at 17.4 and 18.9 $\mu$m, which has a
different spatial distribution than PAHs traced by
the 16.4 $\mu$m feature.

Our imaging spectroscopy demonstrates the spatial separation
between regions of peak PAH emission and
peak C$_{60}$ emission
(Figs. \ref{image1} and \ref{image2}).
\citet{Boersma10} find that the
16.4 $\mu$m feature from different sources correlates with other
PAH features but the 18.9 $\mu$m feature does not.
\citet{Velu08} 
observe the 18.9 $\mu$m feature to peak on the central star
in the reflection nebula 
\object{NGC 2316}
(Parsamian 18).
They also find that the 17.4 and 18.9 $\mu$m features have distinct spatial
distributions in this object,
and argue from this that C$_{60}$ is unlikely to be the carrier of
these features.
Our result that the 17.4 $\mu$m feature is a blend of a PAH feature and
C$_{60}$ in NGC 7023 naturally explains their observations.

\section{Discussion}

To derive the abundance of C$_{60}$, we assume that fullerenes and PAHs 
absorb UV starlight, and then
re-radiate it all as infrared emission features.
We do not include any potential visible fluorescence
by either C$_{60}$ or PAHs. 
For fullerenes, we adopted experimental values of
the absorption cross-section of C$_{60}$ at 0.09--0.35 $\mu$m
\citep{Yasumatsu96, Yagi09}.
The absorption cross-sections from \citet{Yasumatsu96}
and \citet{Yagi09} differ by a factor of 2,
indicating the overall uncertainty in our abundance calculation.
For PAHs, we adopted the absolute absorption cross-section 
at 0.09--0.30 $\mu$m from \citet{LiDraine01}.
We integrated both of these absolute absorption cross-sections
over a 17,000--22,000 K blackbody,
as is appropriate for the central stars of NGC 7023 and NGC 2023.
We find that C$_{60}$ and PAHs have
similar integrated UV absorption strengths per C atom.

We compare the sum of the intensities of the 7.04, 17.4, and 18.9 $\mu$m
features, 
assumed to be due to C$_{60}$,
to the sum of all other infrared emission
features at 5--38 $\mu$m, assumed to be due to PAHs.
We analyze SL and LL spectra with
PAHFIT \citep{Smith07_pahfit}
to find that the ratio of C$_{60}$ to PAH emission is 0.01--0.03 in
our observed positions in NGC 7023 and NGC 2023.
By adopting
a percentage of interstellar carbon in PAHs of 9--18\%,
we derive a percentage of interstellar carbon 
C$_{60}$, $p(C_{60})$, of
0.1--0.6\% in regions of bright
C$_{60}$ emission.

Our $p(C_{60})$ value
is consistent with
other estimates of 
$p(C_{60})$ and of
the percentage of carbon in C$_{60}^+$,
$p(C_{60}^+)$.
\citet{Foing94} estimate 
$p(C_{60}^+)$ = 0.3--0.9\% from two diffuse interstellar
bands at 958 and 963 nm which they attribute to C$_{60}^+$.
\citet{Herbig00} use these same two bands to estimate 
$p(C_{60})^+$ = 0.1--0.3\% in diffuse clouds.  
\citet{Herbig00} places an upper limit of $p(C_{60})$ $<$ 0.0008\% 
in diffuse clouds, showing that C$_{60}$
is primarily ionized in diffuse clouds.
\citet{Moutou99} measure $p(C_{60})$ $<$ 0.3\% 
and $p(C_{60})^+$ $<$ 0.3\% from the lack of
emission features at 7.0--8.5 $\mu$m, at a position in NGC 7023
where the 18.9 $\mu$m C$_{60}$ feature is weak.
\citet{Nucci05} 
find $p(C_{60})$ $<$ 0.6\%, from their
non-detection of the 8.5 $\mu$m feature in absorption towards
R CrB
(\object{HR 5880})
and three massive young stellar objects.
\citet{Cami10} derive $p(C_{60})$ = 1.5\% 
in planetary nebula Tc 1.

The relative intensities of the C$_{60}$  
bands provide information 
on the conditions in which the molecule emits. 
To probe these conditions, we use a Monte Carlo code,
based on a micro-canonical formalism, 
developed to simulate the emission cascade of PAHs following the 
absorption of a UV photon
\citep{Joblin02, Mulas06}.
We calculate the 
evolution of the internal energy in the molecule and the 
number of photons emitted in each infrared band during 
the cooling cascade. 
We calculate the 
emission spectrum after C$_{60}$ absorbs a 
5, 10 or 15 eV photon, reaching temperatures of 
800, 1200 or 1570 K at the beginning of the cooling cascade. 
We use the list of modes from
\citet{Menendez00} and
infrared intensities (at 0 K) 
calculated using density functional theory 
\citep{Choi00, Schettino01}.

We compare the predicted line intensities with the observations
in Table \ref{table1}.
We find the ratio of the intensities of the 17.4 and 18.9 $\mu$m bands is 
not sensitive to the energy of the absorbed UV photon. 
The observed ratio varies but this is likely because
the 17.4 $\mu$m band is blended with a PAH feature.
The ratio of the intensities of the 7.04 and 18.9 $\mu$m 
bands, however, is 
very sensitive to the absorbed UV photon energy. 
Table \ref{table1} shows the 7.04 $\mu$m 
intensity in NGC 7023 is 
consistent with the cooling cascade 
of C$_{60}$ excited by UV 
photons with a mean energy of 10 eV.
A lower photon energy is suggested for
NGC 2023,
perhaps because of its blister geometry 
with the star in front of a dense molecular
cloud.
If C$_{60}$ in
NGC\,2023 is excited by lower energy photons,
then this
would also affect the C$_{60}$ abundance
derived there.
More detailed modeling will be needed
to clarify this.

Our interpretation of the NGC 7023 C$_{60}$ emission
differs from the very recent results of \citet{Cami10} 
who concluded
that C$_{60}$ in Tc 1 emits in solid phase at a temperature
of 330 K.
Their analysis, however, is based on an excitation
diagram to derive
a single emission temperature, which is not
appropriate to describe the
cooling cascade of UV-excited molecules.

\section{Conclusions}

We confirm our identification of 
the 17.4 and 18.9 $\mu$m emission features
in NGC 7023 with C$_{60}$ 
\citep{WUS04, SUW07}
by detecting a predicted emission feature at 7.04 $\pm$ 0.05 $\mu$m.
We also detect 7.04, 17.4, and 18.9 $\mu$m emission features in NGC 2023.
We demonstrate that the 17.4 $\mu$m emission feature
in NGC 7023 is a blend of C$_{60}$ and PAH emission with
{\it Spitzer} imaging spectroscopy.
Our work (\citealt{WUS04}; \citealt{SUW07}; this paper)
is the first firm detection of neutral C$_{60}$ in interstellar space.

We find that the infrared emission in NGC 7023 is consistent with
the emission of gas-phase C$_{60}$ excited by
UV photons. 
Further modeling is 
required to explain the observations of NGC 2023.
The percentage of interstellar carbon in
C$_{60}$
is 0.1--0.6\% in NGC 7023.
This is consistent with previous estimates of, 
and limits on, the interstellar C$_{60}$
and C$_{60}^+$ abundances.

\acknowledgments

We thank Nick Abel, Lou Allamandola, Bruce Draine, 
Alain L\'eger, and Farid Salama for
useful conversations,
Dominique Toublanc for support with the Monte Carlo
code,
and Mike Jura for suggesting the C$_{60}$ identification.
This work is based on observations made with the Spitzer Space Telescope, 
which is operated by the Jet Propulsion Laboratory, 
California Institute of Technology, under a contract with NASA.
Support for this work was provided by NASA 
through an award issued by JPL/Caltech.

{\it Facilities:} \facility{Spitzer (IRS)}.

\end{document}